\documentclass[iop]{emulateapj-rtx4}   
\usepackage{natbib}
\usepackage[backref,breaklinks,colorlinks,citecolor=blue]{hyperref}

\shorttitle{$^{3}$He-rich Solar Energetic Particle Events}
\shortauthors{Nitta et al.}

\begin{document}


\title{Solar Sources of $^{3}$H\lowercase{e}-rich Solar Energetic Particle Events in Solar
  Cycle 24}


\author{Nariaki V. Nitta\altaffilmark{1}, Glenn M. Mason\altaffilmark{2},
  Linghua Wang\altaffilmark{3}, Christina M. S. Cohen\altaffilmark{4}, and Mark E. Wiedenbeck\altaffilmark{5}
}


\altaffiltext{1}{Lockheed Martin Advanced Technology Center,
  Dept/A021S, B/252, 3251 Hanover Street, Palo Alto, CA 94304, USA; \email{nitta@lmsal.com}}
\altaffiltext{2}{Applied Physics Laboratory, Johns Hopkins University,
  Laurel, MD 20723, USA; \email{glenn.mason@jhuapl.edu}}
\altaffiltext{3}{Institute of Space Physics and Applied Technology,
  Peking University, Beijing 100871, China; \email{wanglhwang@gmail.com}}
\altaffiltext{4}{California Institute of Technology, Pasadena, CA
  91125, USA; \email{cohen@srl.caltech.edu}}
\altaffiltext{5}{Jet Propulsion Laboratory, California Institute of
  Technology, Pasadena, CA 91109, USA; \email{mark.e.wiedenbeck@jpl.nasa.gov}}


\begin{abstract}
Using high-cadence extreme-ultraviolet (EUV)
images obtained by the Atmospheric Imaging Assembly (AIA) on board 
the {\it Solar Dynamics Observatory}, we investigate the solar
sources of 26~$^{3}$He-rich solar energetic particle (SEP) events
at $\lesssim$1~MeV~nucleon$^{-1}$ that were well-observed by the {\it
  Advanced Composition Explorer}
during solar cycle 24. 
Identification of the solar sources is based on the association of
$^{3}$He-rich events with type III radio bursts and electron events 
as observed by {\it Wind}.
The source locations are further verified in
EUV images from the 
{\it Solar and Terrestrial Relations Observatory}, which provides
information on solar activities in the regions not visible from the Earth.  
Based on AIA observations, $^{3}$He-rich events are not only
associated with coronal jets as emphasized in solar cycle 23 studies,
but also with more spatially extended eruptions. 
The properties of the $^{3}$He-rich events do not appear to be 
strongly
correlated with those of the source regions.
As in the previous studies, the magnetic connection between the source
region and the observer is not always reproduced adequately by 
the simple potential field source surface model
combined with the Parker spiral.  Instead,
we find a broad longitudinal distribution of the source regions extending well beyond the west
limb, with the longitude deviating significantly from that expected
from the observed solar wind speed.   

\end{abstract}





\section{Introduction}

Solar energetic particle (SEP) events are classified into two
types, corresponding to different origins
\citep{1999SSRv...90..413R,2013SSRv..175...53R}. 
Gradual SEP events, which can be intense enough to be space weather
hazardous, are attributed to shock waves driven by fast and wide
coronal mass ejections (CMEs) as supported by their close correlation
\citep[e.g.][]{1984JGR....89.9683K}.  Another class is often called
``impulsive'' and characterized first by anomalously enriched $^{3}$He and
heavy ions.  They have been known for a long time 
\citep[e.g.][]{1970ApJ...162L.191H}, but the origin of $^{3}$He-rich
SEP events 
is still not well-understood, although theoretical models
on the basis of stochastic acceleration
\citep[e.g.][]{2006ApJ...636..462L}
have been developed.

One of the reasons why $^{3}$He-rich events still lack compelling explanation may be
the difficulty of observing their sources in the corona.  Unlike gradual SEP events,
the association of $^{3}$He-rich events with CMEs is not high 
\citep[][but see below for recent results]{1985ApJ...290..742K}.
Until recently it has been generally believed 
that $^{3}$He-rich events could arise without detectable solar activities
or be associated with minor flares or brightenings at most
\citep{1987SoPh..107..385K,1988ApJ...327..998R}.

During solar cycle 23, new studies have started to reveal the
properties of solar activities that were possibly related to $^{3}$He-rich events,
thanks to the uninterrupted full-disk images of the
solar corona by the {\it Solar and Heliospheric Observatory (SOHO)}.  Using images
from the Extreme-ultraviolet Imaging Telescope 
\citep[EIT;][]{1995SoPh..162..291D}
and Large Angle Spectroscopic Coronagraph 
\citep[LASCO;][]{1995SoPh..162..357B},
\cite{2006ApJ...639..495W} reported on coronal jets (characterized by
linear features)
typically from coronal hole boundaries around the times of 25 $^{3}$He-rich events. 
Some of these jets were seen to extend into the high corona and
observed as narrow CMEs.  Indeed, due largely to the high sensitivity
of LASCO,
the association of $^{3}$He-rich events with CMEs
has become higher than that known earlier if we include these
narrow CMEs \citep{2001ApJ...562..558K}. 

$^{3}$He-rich events are most commonly observed at
$\lesssim$1~MeV~nucleon$^{-1}$.
These particles take several hours to travel to 1~AU, with a
wide temporal spread depending on the effective path length that they
traverse in interplanetary space. 
Therefore it is not straightforward to isolate the solar
activities related to $^{3}$He enrichment.  It has been known that $^{3}$He-rich events
are often associated with type III bursts at $<$2~MHz \citep{1986ApJ...308..902R} and
1\,--\,100~keV electron events \citep{1985ApJ...292..716R}.  Using these two
observables, \cite{2006ApJ...650..438N} identified the solar sources of 69
discrete $^{3}$He-rich events, many of which were jets.  
However, they failed to find compelling solar activities for
$\sim$20\% of $^{3}$He-rich events even when they had good
coverage of full-disk coronal images.

There are three possibilities for $^{3}$He-rich events without solar
activities detectable in EUV and X-ray images. 
First, the acceleration and injection of $^{3}$He ions
may not produce detectable EUV and X-ray emission, as could 
result, e.g., from flare-like processes that occur high up and
leave no traces in the low corona
\citep{1991ApJ...366L..91C}. 
This scenario is similar to the one
that was originally proposed for impulsive
electron events whose power-law spectrum extended down to $\sim$2~keV
\citep{1980ApJ...236L..97P}.
Second, the associated solar activities (such
as jets) may last for too short a time to be detected by the EIT,
which had typically a $\sim$12~m cadence. Lastly, the source region
may be located on the far side and the processes responsible for
$^{3}$He enrichment are limb-occulted.   
  
The primary objective of this paper is to 
explore these possibilities using the new
capabilities that have become available in solar cycle 24.  The Atmospheric Imaging
Assembly \citep[AIA;][]{2012SoPh..275...17L}
on the {\it Solar Dynamics Observatory (SDO)}, which was launched in
February 2010, 
takes full disk images every 12 seconds in a wide range of coronal to
chromospheric temperatures.  This is a significant improvement over
the EIT whose high-rate ($\sim$12~m) data were taken only in one
wavelength.  
With AIA we can detect
minor transient activities that are short-lived and in narrow
temperature ranges.  Furthermore, we now continuously observe the far side of the
Sun as viewed from the Earth, using the EUV Imager 
\citep[EUVI;][]{2004SPIE.5171..111W,2008SSRv..136...67H}
on the {\it Solar and Terrestrial Relations Observatory (STEREO)},
which consists of twin spacecraft that have separated from the Sun-Earth line with the rate of
$\sim$22$\arcdeg$ a year since 2006. 
The EUVI has taken full-disk images with a typical 5~m (10~m) cadence in the 195~\AA\ (304~\AA) channel during the period that
  overlaps with the {\it SDO}.
We can readily determine
whether $^{3}$He-rich events without associated solar activities 
detected in near-Earth observations
are attributable to
activities behind the limb.  

In the next section, we show how the $^{3}$He-rich events are
selected and give a brief overview of their properties.  
Our procedure for finding the solar source is described in \S3, 
using one of the selected events as an example. In \S4, we show the results of the analysis of 
the entire sample of events.  We discuss them in \S5 in terms of 
the previous studies.  A summary of the work is given in \S6.

\begin{figure*}
\includegraphics[scale=.9]{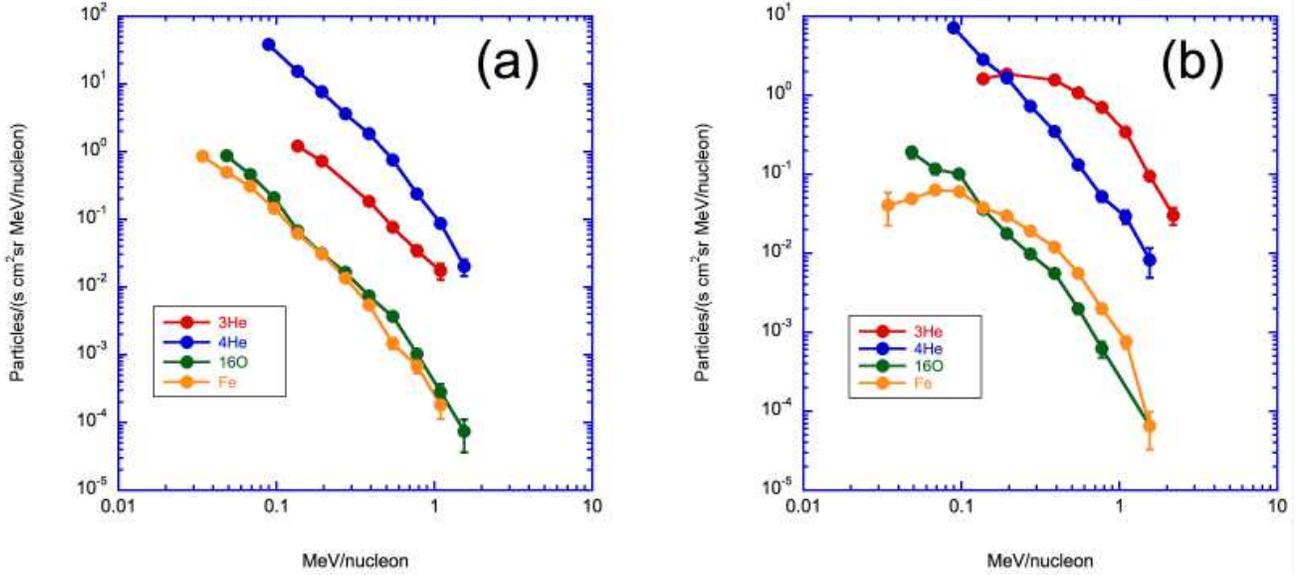}
\caption{Spectra of four ions in (a) Event 24 (2014 May 4) and (b) Event 25 (2014
  May 16).  The $^{3}$He and Fe spectra are a power-law form in panel
  (a), and curved in panel (b).  
}
\end{figure*}



\begin{deluxetable*}{ccccccc}
\tabletypesize{\scriptsize}

\tablecaption{List of $^{3}$He-rich SEP Events with Clear Injections
  and/or Accompanied by Electron Events} 
\tablecolumns{7} \tablewidth{0pt}

\tablehead{
 \colhead{} & \colhead{} & \colhead{Ratio} & \colhead{$>$ factor~2} & \colhead{$^{3}$He} &
\colhead{Observed} & \colhead{Ratio} \\
\colhead{} & \colhead{$^{3}$He-rich} & \colhead{of}
& \colhead{increase in} & \colhead{Spectral} &
\colhead{by} & \colhead{of} \\
\colhead{ID} & \colhead{Period} & \colhead{$^{3}$He/$^{4}$He\tablenotemark{a}} &
\colhead{$^{3}$He/$^{4}$He\tablenotemark{b}} & \colhead{Form\tablenotemark{c}} &
\colhead{SIS?} & \colhead{Fe/O\tablenotemark{d}} \\
}

\startdata 
 1 & 2010 Oct 17 00:02\,--\,Oct 19 18:00\tablenotemark{e} & 0.79$\pm$0.07 & Yes &
 PL & Yes & 1.26$\pm$0.23 \\  
   & 2010 Oct 19 18:00\,--\,Oct 20 18:00 & 2.16$\pm$0.39 & Yes &
  C & Yes  & 1.93$\pm$0.43 \\  
 2 & 2010 Nov 02 00:02\,--\,Nov 03 23:57\tablenotemark{e} & 1.37$\pm$0.10 & Yes &
 C & Yes & 0.91$\pm$0.32 \\  
 3 & 2010 Nov 14 00:02\,--\,Nov 17 12:00 & 0.25$\pm$0.03 & No &
 C &   & 1.22$\pm$0.18 \\  
   & 2010 Nov 17 15:00\,--\,Nov 18 12:00 & 4.25$\pm$0.62 & Yes &
 C &   & 1.34$\pm$0.35 \\  
 4 & 2011 Jan 27 12:00\,--\,Jan 30 12:00\tablenotemark{e} & 0.08$\pm$0.01 & No &
 PL &   & 1.13$\pm$0.17 \\  
 5 & 2011 Jul 07 18:00\,--\,Jul 10 12:00\tablenotemark{e} & 1.68$\pm$0.09 & Yes &
 C  & Yes & 1.11$\pm$0.12 \\  
 6 & 2011 Jul 31 21:00\,--\,Aug 01 18:00 & 0.05$\pm$0.01 & No &
 PL  &  & 2.66$\pm$0.45 \\  
 7 & 2011 Aug 26 00:01\,--\,Aug 28 12:00 & 0.64$\pm$0.05 & Yes &
 C  &  & 1.51$\pm$0.13 \\  
 8 & 2011 Dec 14 12:00\,--\,Dec 15 23:57 & 0.18$\pm$0.02 & Yes &
 PL  & Yes & 1.35$\pm$0.22 \\  
 9 & 2011 Dec 24 18:00\,--\,Dec 25 03:00 & 0.13$\pm$0.01 & Yes &
 PL  & Yes & 1.44$\pm$0.26 \\  
10 & 2012 Jan 03 00:01\,--\,Jan 04 06:00\tablenotemark{e} & 0.08$\pm$0.01 & No &
 PL  &   & 0.99$\pm$0.07 \\  
11 & 2012 Jan 13 12:00\,--\,Jan 14 23:56 & 1.35$\pm$0.14 & Yes &
 PL  & Yes & 2.39$\pm$0.47 \\  
12 & 2012 May 14 12:00\,--\,May 16 18:00 & 0.05$\pm$0.01 & Yes &
 PL  & Yes & 0.40$\pm$0.04 \\  
13 & 2012 Jun 08 03:00\,--\,Jun 09 18:00\tablenotemark{e} & 0.36$\pm$0.03 & Yes &
 PL  & Yes & 2.24$\pm$0.35 \\  
14 & 2012 Jul 03 00:00\,--\,Jul 05 18:00\tablenotemark{e} & 0.09$\pm$0.01 & No &
 PL  & Yes & 1.10$\pm$0.10 \\  
15 & 2012 Nov 18 06:01\,--\,Nov 20 06:00 & 0.55$\pm$0.07 & Yes &
 PL  &   & 0.86$\pm$0.18 \\  
   & 2012 Nov 20 06:01\,--\,Nov 21 03:00\tablenotemark{e} & 6.89$\pm$0.71 & Yes &
 C  &    & 1.06$\pm$0.19 \\  
16 & 2013 May 02 12:00\,--\,May 04 06:00 & 0.21$\pm$0.03 & No &
 PL  &   & 1.04$\pm$0.21 \\  
17 & 2013 Jul 17 00:01\,--\,Jul 18 12:00 & 0.39$\pm$0.03 & Yes &
 C  &   & 0.86$\pm$0.18 \\  
18 & 2013 Dec 24 00:01\,--\,Dec 25 12:00 & 1.30$\pm$0.36 & Yes &
 PL  &   & 2.33$\pm$0.86 \\  
19 & 2014 Jan 01 06:00\,--\,Jan 02 18:00 & 0.26$\pm$0.02 & Yes &
 C   & Yes & 1.83$\pm$0.16 \\  
20 & 2014 Feb 06 00:01\,--\,Feb 07 12:00 & 0.49$\pm$0.05 & Yes &
 C   &   & 1.14$\pm$0.12 \\  
21 & 2014 Mar 29 00:01\,--\,Mar 30 23:56 & 0.17$\pm$0.02 & Yes &
 C   &   & 0.36$\pm$0.07 \\  
22 & 2014 Apr 17 22:20\,--\,Apr 19 12:00 & 0.58$\pm$0.04 & No &
 PL  &   & 1.37$\pm$0.12 \\  
23 & 2014 Apr 24 00:01\,--\,Apr 25 23:58 & 1.28$\pm$0.07 & Yes &
 C   & Yes  & 0.71$\pm$0.09 \\  
24 & 2014 May 04 09:00\,--\,May 05 18:00 & 0.14$\pm$0.02 & Yes &
 PL   &   & 0.74$\pm$0.11 \\  
25 & 2014 May 16 06:00\,--\,May 17 18:00 & 14.88$\pm$1.36 & Yes &
 C   & Yes  & 2.13$\pm$0.27 \\  
26 & 2014 May 29 03:00\,--\,May 30 03:00 & 2.64$\pm$0.45 & Yes &
 C   &    & 2.95$\pm$0.40 \\  

\enddata

\tablecomments{
\tablenotetext{a}{In the range of 0.5\,--\,2.0~MeV nucleon$^{-1}$.}
\tablenotetext{b}{In the range of 0.1\,--\,1.0~MeV nucleon$^{-1}$.}
\tablenotetext{c}{In the range of 0.1\,--\,1.0~MeV nucleon$^{-1}$.  Power-Law (PL) or Curved (C).}
\tablenotetext{d}{In the range of 320\,--\,450~keV nucleon$^{-1}$.}
\tablenotetext{e}{Single $^{3}$He-enriched period for multiple type
  III bursts/electron events.}}

\end{deluxetable*}


\section{Event Selection}

The $^{3}$He-rich SEP events that we study in this paper are based on data from the Ultra
Low Energy Isotope Spectrometer \citep[ULEIS;][]{1998SSRv...86..409M}
on board
the {\it Advanced Composition Explorer (ACE)}.  The ULEIS is a
time-of-flight mass spectrometer with geometric factor of
$\sim$1~cm$^{2}$.  We surveyed ULEIS data from May 2010 through May
2014, and selected 26 $^{3}$He-rich events that showed either clear ion injections
or clear $^{3}$He presence preceded by a $>$40~keV electron event that was
detected by the Electron, Proton, and Alpha
Monitor \citep{1998SSRv...86..541G} on {\it ACE}. 
We also chose relatively intense
events in terms of $^{3}$He yields so that we can derive the $^{3}$He spectrum in comparison with
other ions, which may be important for identifying the right 
acceleration mechanisms \citep{2002ApJ...574.1039M}.

In Table~1, we list the selected $^{3}$He-rich events with basic information including the
$^{3}$He/$^{4}$He and Fe/O ratios.  In the second column, the $^{3}$He-rich
period is shown.  Several events lasted longer than two days,
comparable to the multiday events studied by
\cite{2008ApJS..176..497K}.  
These long-lasting $^{3}$He-rich periods may represent either
continuous injections or smeared discrete injections.
During the survey period we dropped 
some multiday $^{3}$He-rich events because they neither have clear
ion injections nor an electron event.
We subdivide the period of a long-duration
event only when more than one injection are clearly identified.
The third column is the $^{3}$He/$^{4}$He ratio
in the 0.5\,--\,2.0~MeV nucleon$^{-1}$ range, which shows a wide
variation from 0.05 to 15.
In the next column we show whether more than a factor of two increase in  
the $^{3}$He/$^{4}$He ratio is observed, measuring how strongly 
the $^{3}$He increase is correlated with $^{4}$He.  
Most events have this attribute.
The $^{3}$He spectral shape in the range of 0.1\,--\,1.0~MeV nucleon$^{-1}$ is
given in the fifth column.
It is nearly evenly distributed between power-law (PL) and curved (C) spectra.
Figure~1 shows examples of the two types.
The sixth column shows whether the $^{3}$He-rich event was observed at 
higher energies ($>$4.5~MeV~nucleon$^{-1}$) by the Solar Isotope Spectrometer 
\citep[SIS;][]{1998SSRv...86..357S} on ACE.  About one half of our events
were also SIS events.
The last column shows the Fe/O ratio, which ranges from 0.4 to 3, 
confirming that our $^{3}$He-rich events are also Fe-rich
\citep{1975ApJ...201L..95H,1986ApJ...303..849M}.




\section{Analysis \,--\, Finding Solar Sources of $^{3}$He-rich Events}

\begin{figure}
\includegraphics[scale=.45]{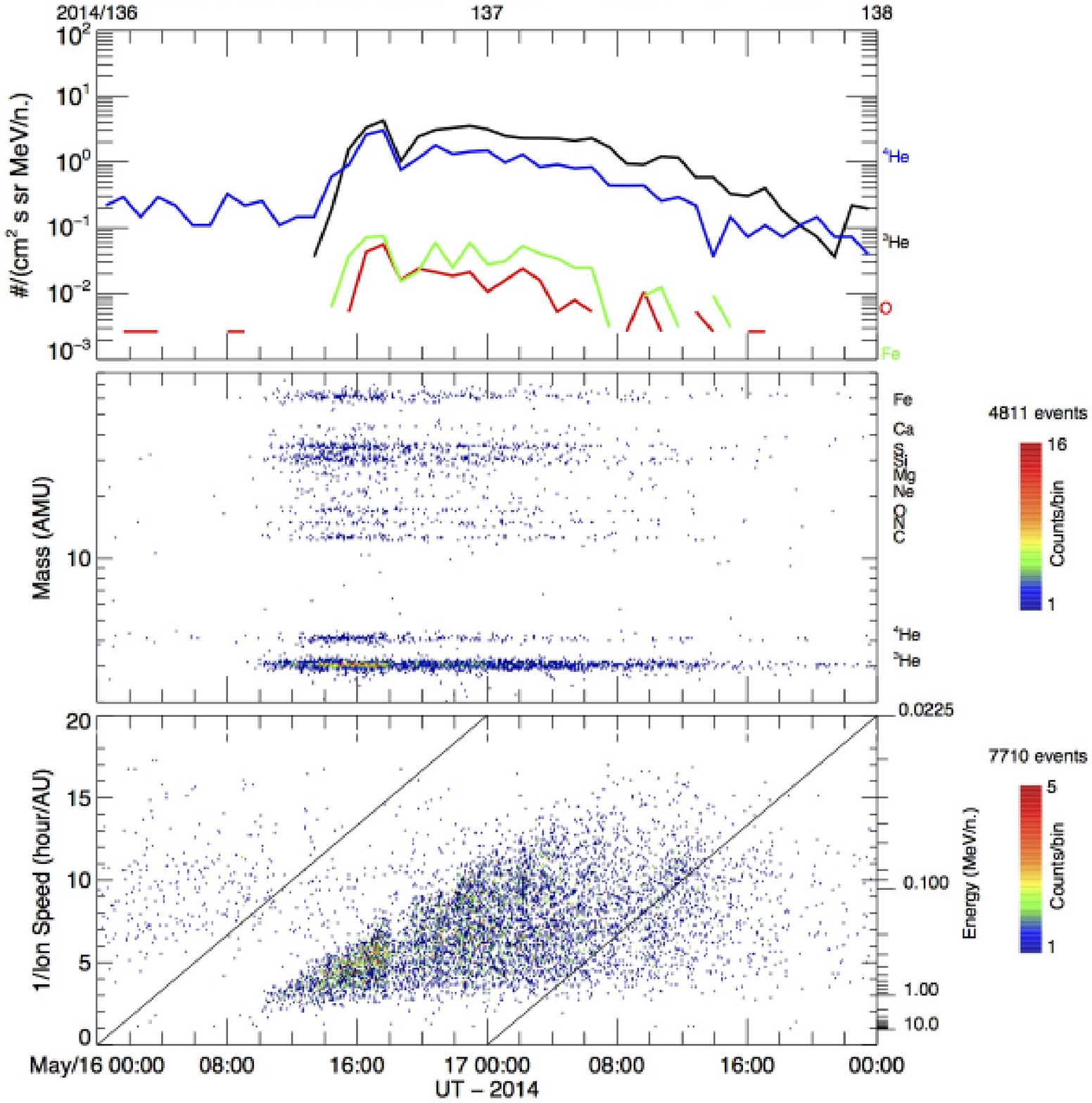}
\caption{$^{3}$He-rich SEP event that started on 2014 May 16.  (a)
  Time profiles of $^{4}$He, $^{3}$He, O and Fe ions in
      0.23\,--\,0.32~MeV nucleon$^{-1}$.  (b) Mass spectrogram for
      elements from He to Fe for ions with energies
      0.4\,--\,10 MeV~nucleon$^{-1}$. (c) Plot of 1/ion speed vs time
      of arrival for ions whose mass ranges are 10\,--\,70~AMU. The two
    lines in cyan show the time range for Figure~3. The two oblique
    lines indicate the arrival times assuming the path length of 1.2~AU.}
\end{figure}

In this work we follow almost 
the same technique as
\cite{2006ApJ...650..438N}, who
connected $^{3}$He-rich events in solar cycle 23 
with solar activities using type III bursts and
electron events.  
We use Event 25 in Table~1 to illustrate how we identify the solar
sources of $^{3}$He-rich events.  This event has the highest 
$^{3}$He/$^{4}$He ratio in our sample (Table~1).  The $^{3}$He and
Fe spectra are characteristically curved in the
0.1\,--\,1~MeV~nucleon$^{-1}$ range (see Figure~1(b)).   
We first find when $^{3}$He ions started to increase significantly.
Figure~2
is a ULEIS multi-panel plot, which consists of
(a) hourly average intensities for  
$^{3}$He, $^{4}$He, O and Fe ions for the energy range 0.23\,--\,0.32~MeV
nucleon$^{-1}$, 
(b) individual ion masses and arrival times for ions of energy 0.4\,--\,10~MeV~nucleon$^{-1}$,
including ions from He to Fe and indicating clear separation of $^{3}$He from $^{4}$He, and 
(c) individual ion reciprocal speed (1/$v$) and arrival time for
ions with mass of 10\,--\,70~AMU. 
Four-day plots in this format are available at the ACE Science
Center
site\footnote{\url{http://www.srl.caltech.edu/ACE/ASC/DATA/level3/index.html}}.  

In this event, $^{3}$He ions started to increase around 10:00~UT on
2014 May 16 (Figure~2(b)).  
We also note velocity dispersion (Figure~2(c)), which could be used to determine
the particle injection time and path length, assuming that (i) the
first-arriving particles of all energies are injected at the same time and that (ii) they
are scatter-free en route to 1~AU.  The upper envelope of the apparent
dispersion indicates the particle injection around 05~UT on May 16 .  
The calculated path length is $\sim$1.8~AU, considerably longer than
1.2~AU, which is indicated by the two oblique lines.  

In this work, we do not utilize the information from ion velocity dispersion.
First, the uncertainty
in the injection time may be more than a few hours.
Second, neither of the assumptions (i) and (ii) may be generally valid.
Furthermore, only four other $^{3}$He-rich periods listed in Table~1
have the velocity dispersion as clear as this one.
We instead take a more conservative approach to find the solar activity that possibly accounts
for the $^{3}$He-rich event.  That is, we 
use a type III burst and electron event in a
long enough time window before the $^{3}$He onset 
in such a way that the presence of an electron event gives preference
to the associated type III burst if there are more than one of
them \citep[see][]{2006ApJ...650..438N}. Then we can determine the
time of the solar activity related to the $^{3}$He-rich event to 
the accuracy of one minute. 
We set
a window 1\,--\,7~hours prior to the $^{3}$He onset as indicated by the two lines in cyan.  
In this event, no major type III bursts are seen if we expand the
window by five more hours (i.e., start of the window at 22:00~UT on
May 15).
Longer time windows are used for a majority of events without clear velocity dispersions.

\begin{figure}
\centerline{\includegraphics[scale=.45]{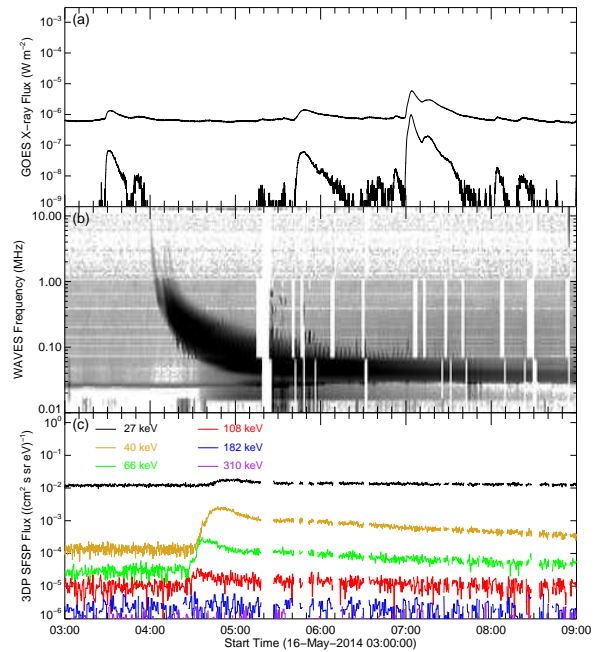}}
\caption{Six-hour period preceding the $^{3}$He-rich event shown in Figure~2
(a) {\it GOES} 1\,--\,8~\AA\ and 0.5\,--4~\AA\ X-ray light curves. (b)
  Radio dynamic spectrum from WAVES.  (c) Electron flux at 1 AU as
  observed by the solid state telescope of 3DP.}
\end{figure}

In Figure~3 we show {\it GOES} soft X-ray light curves, radio dynamic spectrum
and electron time profiles for the time window indicated in Figure~2.
The radio dynamic spectrum below 14~MHz is obtained by the Radio and Plasma Wave
Experiment \citep[WAVES;][]{1995SSRv...71..231B}, 
and the electron time profiles from the Three-dimensional Plasma and
Energetic Particles instrument \citep[3DP;][]{1995SSRv...71..125L}, 
both on the {\it Wind} spacecraft.
Although Figure~3(c) is limited to electrons above $\sim$30~keV as
detected by the solid-state telescope (SST), the 3DP consists also of the
electron electrostatic analyzer (EESA) that observes 3~eV\,--\,30~keV
electrons.  The combined SST and EESA data sometimes produce a clear velocity dispersion over a
wide energy range \citep{2006GeoRL..33.3106W,2012ApJ...759...69W},
which may be used to argue for the scatter-free nature of electron events.
Although the EESA data did not rise above background
in the interval of Figure~3
and therefore are not shown, their availability is one of the reasons we
use 3DP data in this work.

In Figure~3(b) we find type III bursts between
04:00\,--\,04:16~UT.
Figure~3(c) shows an electron event with velocity dispersion.
It is delayed with respect to the start time of the type III by $\sim$15 minutes 
at the highest-energy channel (108~keV)
that observes it.  Figure~3(a) shows a C1.3\footnote{The peak flux
  $I_{peak}$ of 1.3$\cdot 10^{-6}$~W~m$^{-2}$ in the 1\,--\,8~\AA\ channel of the GOES
X-ray Spectrometer.  A C-class flare has $I_{peak}$ of
10$^{-6}$$\leq$$I_{peak}$$<$10$^{-5}$ (W~m$^{-2}$).  
The M-class (B-class) is an order of magnitude higher (lower).} 
flare that peaks at 03:32 UT,
followed by a smaller flare that is better seen in the
0.5\,--\,4~\AA\ channel.  At first, we do not discard the possibility
that these flares may be related to the type III bursts even though
they are too widely separated in time.  
We note that the NOAA event list shows the C1.3
flare coming from AR~12053, which is clearly wrong because the region 
was already 20$\arcdeg$ behind the
west limb.  Therefore
we need to examine full-disk images to
locate them.  AIA images in the 94~\AA\ channel
that has the peak temperature response around 7\,--\,8~MK show that they come
primarily from AR~12063 (N10 E28) and secondarily from AR~12057 (N17
W40). Despite the effort to locate the minor flares, however, we find
a distinctly new pattern after 04:00~UT, much closer in time to the type III
bursts, which is a jet-like ejection in a quiescent region at S12 W44.
This is clearly seen in AIA images in multiple
channels for the next 15 minutes.
We therefore consider this jet to be associated with the type III
bursts and the $^{3}$He-rich event.  

\begin{figure}
\includegraphics[scale=.45]{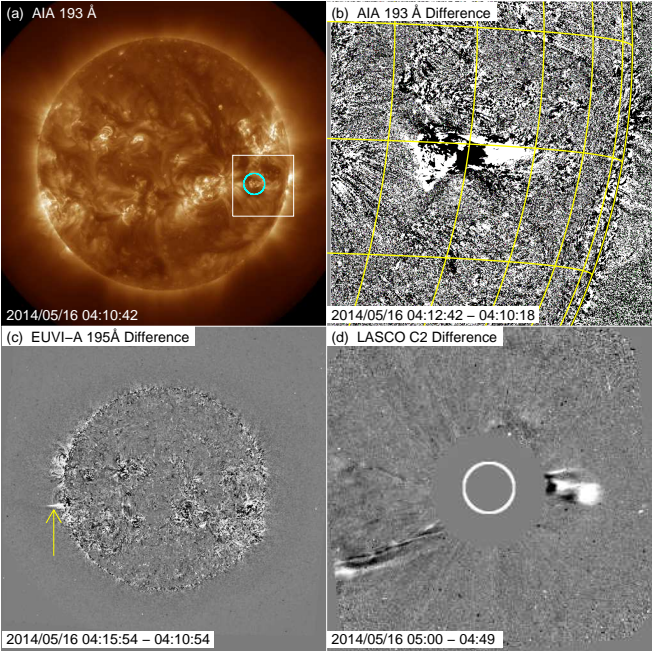}
\caption{(a) AIA 193~\AA\ image around the time of the type III burst
  associated with the $^{3}$He-rich event in Figure~2.  The area encircled
  in cyan contains a jet.  The box shows the field of view of the
  image shown in (b).  (b) AIA 193~\AA\ difference image
  expanded in a limited field of view.  The grid of heliographic
  longitude and latitude is overlaid in yellow at a spacing of
  10$\arcdeg$.  The leg of the jet is located around W44. 
(c) Difference
  image in the 195~\AA\ channel around the same time as taken by the
  EUVI on {\it STEREO-A}. The arrow in yellow points to the same jet.
  (d) LASCO C2 difference image taken shortly after the jet,
  confirming a narrow CME.}
\end{figure}

Figure~4(a) shows an AIA 193~\AA\ full-disk image on which the
location of the jet is encircled.  An enlarged view of the jet 
is shown as a difference image in
Figure~4(b).  Its field-of-view is marked in Figure~4(a).
The jet is extended both in time
and space, and it would probably have been detected by EIT.
As in previous studies
\citep{2006ApJ...639..495W,2006ApJ...650..438N,2008ApJ...675L.125N},
this jet is later identified as a relatively narrow CME (Figure~4(d)). 
The same jet also appears on the east limb in a {\it STEREO-A} EUVI
difference image (Figure~4(c)).  The footpoint of the
jet should be at longitude of 
E114 as viewed from {\it STEREO-A}, 
which was then located at 158$\arcdeg$ west of the
Sun-Earth line.  In other events, the coronal signatures can be much
weaker in the 193~\AA\ channel than those in this event.  Therefore we examine images in
different channels and various forms, that is, intensity, running
difference and ratio, and base (pre-event) difference and ratio, applying wide ranges of
normalization factors.  

\begin{figure}
\includegraphics[scale=.45]{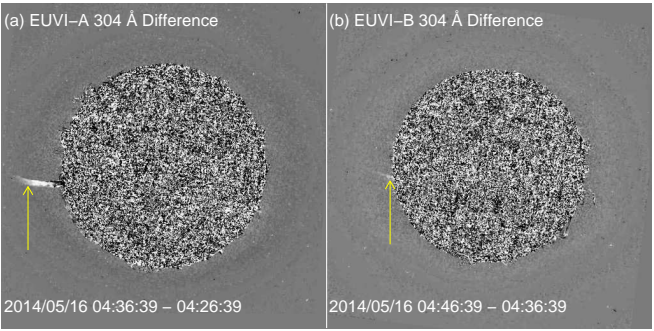}
\caption{Difference images in the 304~\AA\ channel of (a)
  EUVI-A and (b) EUVI-B, showing the jet extended to higher altitudes
  as indicated by the arrows.}
\end{figure}

The lower part of this jet was more clearly observed later
in 304~\AA\ images (Figure~5) that primarily
represent chromospheric temperatures.  It is
interesting to note that the jet is seen even from {\it
  STEREO-B}, for which the region's longitude was E151 (see Figure~6).  
The corresponding occultation height is 1.08~R$_{\sun}$ from the photosphere.
This example demonstrates that {\it STEREO} data are not only useful to confirm
the locations of the solar activities associated with $^{3}$He-rich
events, but also to determine their extensions in height.

\begin{figure}
\centerline{\includegraphics[scale=.38]{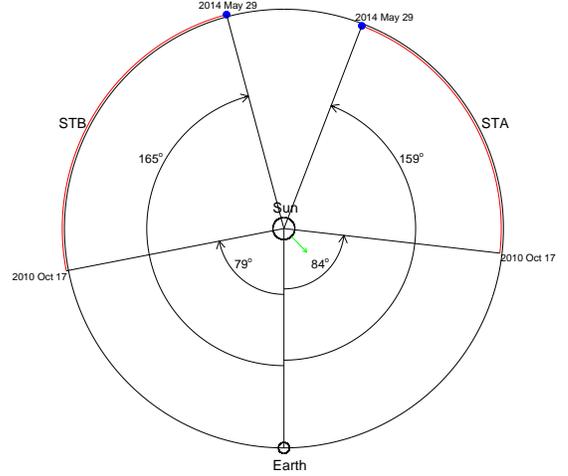}}
\caption{{\it STEREO} angular separation (indicated in red) during the period our
  $^{3}$He-rich events occurred (2010 October 17\,--\,2014 May 29).
  The blue dots and green arrow show, respectively, the positions
  of {\it STEREO} and the source region for Event 25 (2014 May 16).
 }
\end{figure}

To further illustrate the unique information coming from {\it STEREO} observations, 
Figure~6 shows the locations of {\it STEREO-A (STA)} and {\it STEREO-B
  (STB)} during the survey period
for the $^{3}$He-rich events studied in this paper.  EUVI
on {\it STEREO-A} unambiguously observed as disk events
those that were behind the west limb as seen from Earth.  Early in the
period, events near central meridian in Earth view were observed as limb events by
both {\it STEREO}.  Toward the end of the period, {\it STEREO-B}
observed western-hemisphere events as occulted by the east limb
as was the case for Figure~5(b).

\begin{figure*}
\includegraphics[scale=.90]{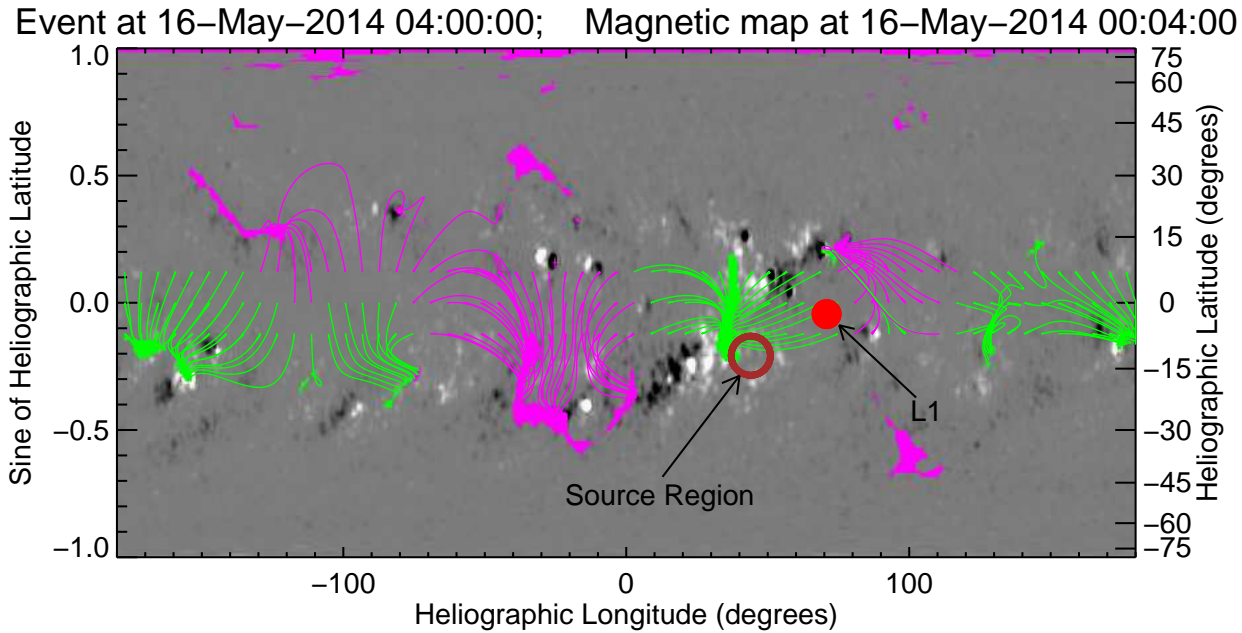}
\caption{Synoptic magnetic map shortly before the $^{3}$He-rich SEP
Event 25.  The Carrington longitudes are translated to the
  Earth-view longitudes.  The circles indicate the location 
  of the solar activity, represented as Source Region, on
  the photosphere and the magnetic connection of L1 to the source
  surface. Open field regions are marked in two colors (green:
  positive, and pink: negative).  Open field lines reaching the
  source surface at the ecliptic and $\pm$7$\arcdeg$ latitudes are
  indicated in the colors matching the polarities. }
\end{figure*}

Now we study the source region of the $^{3}$He-rich event in terms of
the global magnetic field that becomes part of solar wind sampled at
1~AU.  In particular, we are interested in the relation of the source
region with the open field lines that are connected to {\it ACE},
because the observed particles propagate along them.
In Figure~7
we show the relation of the source region with  
the magnetic footprint of the observer at {\it ACE}, 
marked L1, in terms of computed open field lines.
The grayscale image is a synoptic magnetic map, 
not in the usual Carrington coordinates, but in the Stonyhurst
(Earth-view) coordinate systems, in which zero of the 
X-axis corresponds to zero heliographic longitude.  
The source region is plotted in its photospheric location, i.e. S12 W44,
whereas the L1 symbol is in the projected location on the source surface 
(at 2.5~R$_{\sun}$ from the Sun center) of the foot-point of the field
line that crosses L1.  In order to obtain the latter, we assume the
Parker spiral that corresponds to the observed solar wind speed averaged over 5
hours around the type III burst.  The magnetic field is extrapolated
from the photosphere, using the potential field source surface
(PFSS) model.

Specifically, we base this study on the PFSS package 
in SolarSoft as implemented 
by \cite{2003SoPh..212..165S}, which includes updates of synoptic
magnetic maps every 6 hours.  
They have equal-area pixels with the resolution of 1~deg$^{2}$ at
  the equator.
These maps are used as the lower
boundary conditions for the PFSS extrapolation.  They are constructed
from (1) the longitudinal magnetograms taken by 
the Helioseismic and Magnetic Imager 
\citep[HMI;][]{2012SoPh..275..207S,2012SoPh..275..229S}
for the area 60$\arcdeg$ from disk center, 
and (2) a flux transport model \citep{2001ApJ...547..475S} for the
remainder of the solar surface.  Here we overplot
coronal holes or contiguous open field regions as
filled areas.  They are color-coded depending on the polarity of
the photospheric footpoint (green for positive and pink for negative).
We use the same color code to show open field lines  
that reach the source surface at the ecliptic and
$\pm$7$\arcdeg$ latitudes.  We include the latter field lines to show 
the latitudinal uncertainties of the PFSS results
\citep[cf.][]{2008ApJ...673L.207N}.

In Event 25, the photospheric location of the jet is close to the
coronal hole of positive polarity, and to open field lines that are
connected to L1.  This is discussed more quantitatively in the next
section.  The positive polarity of these open field lines is consistent with
the observed polarity of the Interplanetary Magnetic Field (IMF) 
during the whole period of the $^{3}$He-rich
event, and also with that indicated by the anisotropy of electrons 
observed in 3DP pitch-angle data.  In this work, we examine solar wind
data from {\it Wind} and {\it ACE}, not only to compare the polarities
of the IMF and the source regions of the $^{3}$He-rich
events, but also to identify interplanetary CMEs (ICMEs), corotating
interaction regions and other non-steady structures that may modify
the Sun-Earth magnetic field connections.







\section{Results}

\begin{deluxetable*}{clccccccrrccrrcrc}
\tabletypesize{\scriptsize}

\tablecaption{Various Observations Related to $^{3}$He-rich SEP Events} 
\tablecolumns{17} \tablewidth{0pt}

\tablehead{
\colhead{} & \colhead{} & \colhead{} & \multicolumn{3}{c}{Onset Times\tablenotemark{a}} & 
\colhead{} & \multicolumn{4}{c}{Source (Flare)} & \colhead{} & 
\multicolumn{2}{c}{CME} & \colhead{} &  \multicolumn{2}{c}{Solar Wind} \\
\cline{4-6} \cline{8-11} \cline{13-14} \cline{16-17} \\
\colhead{ID} & \colhead{$^{3}$He Start} & \colhead{} & \colhead{Type III} & 
\colhead{Electron} &  \colhead{Flare} & \colhead{} & \colhead{GOES} & 
\colhead{Loc.} &  \colhead{Pol.\tablenotemark{b}} &  \colhead{Mot.\tablenotemark{d}} &   \colhead{}  &
\colhead{AW\tablenotemark{e}} & \colhead{$v_{pos}$\tablenotemark{f}} & \colhead{} & \colhead{$v_{p}$\tablenotemark{g}} &
\colhead{Pol.\tablenotemark{h}} \\
\colhead{} & \colhead{} & \colhead{} & \colhead{} & 
\colhead{} &  \colhead{} & \colhead{} & \colhead{} & 
\colhead{} &  \colhead{PFSS\tablenotemark{c}} &  \colhead{} &   \colhead{}  &
\colhead{} & \colhead{} & \colhead{} & \colhead{} &
\colhead{} \\
 }

\startdata 
 1 & 2010 Oct 17   & & Oct 17 08:55 & 09:10 & 08:52 & & C1.7 & S18 W33  &
C~2 & E & & 54 & 304 & & 383 & $+$ \\  
   &             & & Oct 19 06:48 & 07:05 & 06:45 & & C1.3 & S18 W57  &
C~4 & E & & 77 & 385 & & 406 & $-$ \\  
 2 & 2010 Nov 02 & & Nov 02 07:28 & 07:55 & 07:26 & & B1.9 & N20 W90  &
$-$~2 & L & & 67 & 253 & & 337 & $+$ \\  
 3 & 2010 Nov 14 & & Nov 13 23:52 & N & 23:50 & & C1.1 & S23 W26  &
$-$~1 & E & & 63 & 442 & & 465 & $-$ \\  
   &             & & Nov 17 08:09 & 08:20 & 08:07 & & B3.4 & S22 W72  &
C~3 & J & & 41 & 639 & & 498 & $-$ \\  
 4 & 2011 Jan 27 & & Jan 27 08:45 & 08:40 & 08:40 & & B6.6 & N14 W80  &
$-$~1 & L & & 52 & 316 & & 304 & $-$ \\  
 5 & 2011 Jul 07 & & Jul 07 14:27 & 14:50 & 14:25 & & B7.6 & N15 W91  &
$-$~1 & J & & 33 & 715 & & 348 & $-$ \\  
 6 & 2011 Jul 31 & & Jul 31 19:01 & 19:15 & 19:01 & & C1.7 & N18 W51  &
$-$~1 & L & & 47 & 280 & & 659 & $-$ \\  
 7 & 2011 Aug 26 & & Aug 26 00:42 & 01:05 & 00:41 & & B4.4 & N18 W62  &
$-$~1 & J & & 26 & 305 & & 422 & $-$ \\  
 8 & 2011 Dec 14 & & Dec 14 03:11 & 03:45 & 03:01 & & C3.5 & S18 W86  &
C~3 & E & & 40 & 576 & & 441 & $-$ \\  
 9 & 2011 Dec 24 & & Dec 24 11:58 & 12:15 & 11:20 & & C4.9 & N16 W92  &
$-$~1 & E & & 42 & 536 & & 352 & $-$ \\  
10 & 2012 Jan 03 & & Jan 03 01:51 & 02:10 & 01:40 & & B5.0 & S20 W63  &
$+$~1 & J & & 29 & 670 & & 411 & $+$ \\  
11 & 2012 Jan 13 & & Jan 13 09:08 & 09:20 & 09:10 & & N & N16 W111  &
$-$~4 & J & & 62 & 350 & & 501 & $+$ \\  
12 & 2012 May 14 & & May 14 09:35 & 09:55 & 09:35 & & C2.5 & N08 W45  &
$-$~1 & E & & 48 & 551 & & 438 & $-$ \\  
13 & 2012 Jun 08 & & Jun 08 07:15 & 07:45 & 07:11 & & C4.8 &  N13 W40  &
$-$~1 & L & & 34 & 308 & & 593 & $-$ \\  
14 & 2012 Jul 03 & & Jul 02 18:04 & 18:45 & 18:03 & &  C4.5 &  N16 W09  &
$-$~1 & J & & N & N & & 638 & $-$ \\  
15 & 2012 Nov 18 & & Nov 18 04:00 & 04:30 & 03:55 & &  C5.7 &  N08 W07  &
C~4 & E & & 29\tablenotemark{i} & 49\tablenotemark{i} & & 407 & $+$ \\  
  &              & & Nov 20 01:30 & N & 01:31 & &  N &  S17 W60  &
C~2 & J & & N & N & & 390 & $+$ \\  
16 & 2013 May 02 & & May 02 04:55 & 05:25 & 04:58 & &  M1.1 &  N10 W25  &
$-$~1 & E & & 99 & 671 & & 460 & $-$ \\  
17 & 2013 Jul 17 & & Jul 16 20:21 & 20:35 & 20:18 & &  B1.8 &  N21 W70  &
C~4 & J & & 21 & 270 & & 349 & $+$ \\  
18 & 2013 Dec 24 & & Dec 24 12:45 & 20:35 & 12:42 & & C1.2 &  S17 W92  &
$+$~1 & J & & 21 & 270 & & 285 & $+$ \\  
19 & 2014 Jan 01 & & Jan 01 07:24 & 07:45 & 07:21 & & C3.2 &  S13 W47  &
$+$~1 & J & & 185\tablenotemark{j} & 465\tablenotemark{j} & & 375 & $+$ \\  
20 & 2014 Feb 06 & & Feb 05 22:55 & 23:10 & 22:55 & & N &  S13 W84  &
C~3 & J & & N & N & & 385 & $+$ \\  
21 & 2014 Mar 29 & & Mar 28 21:00 & 21:15 & 20:57 & & C1.0 &  S17 W66  &
C~3 & J & & 10\tablenotemark{i} & 314\tablenotemark{i} & & 427 & $-$ \\  
22 & 2014 Apr 17 & & Apr 17 21:58 & 22:20 & 21:50 & & C3.2 &  S15 W24  &
$+$~1 & E & & 119\tablenotemark{j} & 824\tablenotemark{j} & & 388 & $+$ \\  
23 & 2014 Apr 24 & & Apr 24 00:40\tablenotemark{k} & 01:35 & 00:50 & & N &  S18 W102  &
C~2 & E & & 90 & 601 & & 414 & $-$ \\  
24 & 2014 May 03 & & May 03 20:16 & 21:00 & 20:18 & & C1.8 & S11 W36  &
$+$~1 & E & & 218\tablenotemark{j} & 494\tablenotemark{j} & & 353 & $-$ \\  
25 & 2014 May 16 & & May 16 03:57 & 04:25 & 20:18 & & N & S12 W44  &
$+$~1 & J & & 43 & 592 & & 327 & $+$ \\  
26 & 2014 May 29 & & May 29 02:37\tablenotemark{k} & 03:00 & 02:50 & & N & S14 W105  &
$+$~1 & E & & 35 & 418 & & 330 & $+$ \\  

\enddata

\tablecomments{
\tablenotetext{a}{Time referred to 1~AU.}
\tablenotetext{b}{Polarity of the source region, `-', `+' or `C'.
  `C' stands for `closed', meaning that the location has no open field lines within
  20$\arcdeg$ heliographic coordinates.}
\tablenotetext{c}{Measure of PFSS extrapolations, 1: best to 4: worst
  (see text).}
\tablenotetext{d}{Characteristic motions seen in EUV images.  J: jet, E: eruption, L: Large-scale coronal
  propagating front (EUV wave).} 
\tablenotetext{e}{CME angular width in degrees. }
\tablenotetext{f}{CME plane-of-the-sky speed in km~s$^{-1}$.}
\tablenotetext{g}{Solar wind speed (protons) in km~s$^{-1}$.}
\tablenotetext{h}{Polarity of solar wind magnetic field during the
  electron and $^{3}$He-rich events.}
\tablenotetext{i}{Derived using {\it STEREO} COR-1 data.}
\tablenotetext{j}{Not clear if the CME consists of a single eruption.}
\tablenotetext{k}{Time from {\it STEREO} Waves \citep{2008SSRv..136..487B}.}
}
\end{deluxetable*}

We perform the analysis described in the previous section for all
the selected  $^{3}$He-rich events.
Table~2 gives the results.
First of all, 
we always found at least one type III burst in a time window up to 10 hours
before the observed $^{3}$He onset, confirming the excellent
correlation between the two phenomena
\citep[see][]{2006ApJ...650..438N}.  Moreover, except for a few cases,
there is no ambiguity finding the type III burst that appears to be
associated with the $^{3}$He-rich event.  Even when multiple type III
bursts are present within the time window 
preceding a $^{3}$He-rich event (as identified by the subscript e
  in Table~1), 
they often share the same
source region \citep[e.g. Event 5, see][]{2014ApJ...786...71B}. 
It is possible that 
they are all associated with the $^{3}$He-rich event.  
In such cases, Table~2 lists only a representative type III burst for
each  $^{3}$He-rich interval in Table~1.

Almost all 
the type III bursts accompany an electron event as in the fourth
column, which may appear contradictory to
\cite{2006ApJ...650..438N}, who found that $^{3}$He-rich events are much
less frequently associated with electron events than with type III
bursts (62\% vs 95\%.)  However, this largely results from one
of our selection criteria.  Note that not all the electron events
exhibit the velocity dispersion that points to a release time close to
the type III burst.
Some of them start long
(e.g. $>$30~m) after the type III burst or flare,
have gradual time profiles, or show no velocity dispersion.  
Although these properties may suggest their different origins,
it is beyond the scope of this work to study electron events in detail.


\begin{figure*}
\includegraphics[scale=.90]{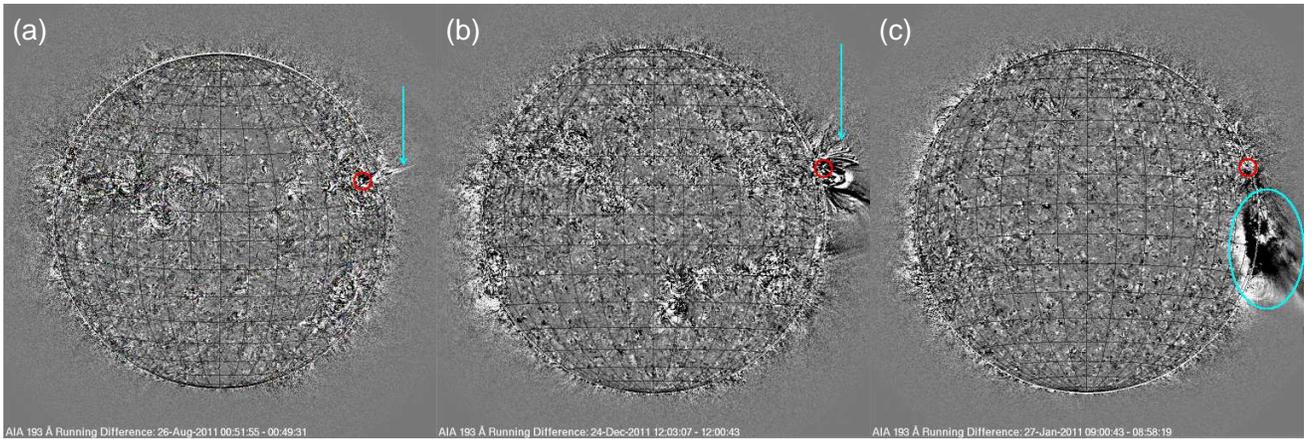}
\caption{Three types of motions in EUV images 
  associated with $^{3}$He-rich events. (a) Jet (in Event 7), (b)
  Eruption (in Event 9), and (c) Large-scale propagating front (LCPF,
  Event 4).}
\end{figure*}

The magnitude
of the associated flare is given in the sixth column.  
Most events are associated with flares that are 
the {\it GOES} C-class 
or below.  Some of the type III bursts have no associated {\it GOES} flare, 
labeled N (no flare).  Events in this category tend to have
high $^{3}$He/$^{4}$He ratios as shown in Table~1 (e.g. Event 25 featured in \S~3), which
is consistent with earlier findings \citep[see][]{1988ApJ...327..998R}.  
They also include some events with curved $^{3}$He spectrum.

The ninth column shows the characteristic motion of the solar
activity possibly responsible for $^{3}$He-rich events.  It is one of the three
types (see Figure~8). 
If it contains linear features (see Figure~8(a)), we
label it a jet (J).  If it shows larger angular
expanse, probably involving closed loop structures like CMEs that are not
narrow (Figure~8(b)),
it is labeled an eruption (E).  The distinction between these two can be
subjective and dependent on projection and observed temperatures.  
Lastly the motion may reach 
large distances (Figure~8(c)), which we classify as 
the EIT wave or large-scale coronal propagating front  
\citep[LCPF, see][]{2013ApJ...776...58N}.  We label such
an event L.   
We note that most events belong to either E or J with only a few 
showing large-scale motions.  There is no clear correlation of
these motions
with the basic properties of $^{3}$He-rich events in Table~1,
except that events with high $^{3}$He/$^{4}$He ratios and curved $^{3}$He spectra are more often associated with jets.

Now we look at the association of $^{3}$He-rich events with CMEs
whose angular width and velocity (in the plane of the sky) 
are given in the tenth and eleventh columns.  
They are mostly taken from the CDAW
catalog\footnote{\url{http://cdaw.gsfc.nasa.gov/CME\_list/}}, but we
also made independent identification and measurement for unclear
cases.  
In addition, we examined data from the {\it STEREO} COR-1 and
  COR-2 coronagraphs \citep{2008SSRv..136...67H} when no CMEs were
  found in LASCO data (e.g. Events 15 and 21).
At first we expect CMEs to be influenced by the three types of
coronal motions. For example, CMEs associated with an eruption
may be wider than those associated with a jet.  The same may be true
for CMEs associated with a LCPF as compared with those that are associated
with an eruption or a jet.  Such expectations are not supported by the
observed CME width.  The CME velocity is also uncorrelated with
the coronal motions.
In short, we still do not understand the relation of $^{3}$He-rich events with CMEs,
which often, but not always, accompany them and whose properties vary.

\begin{figure}
\includegraphics[scale=.45]{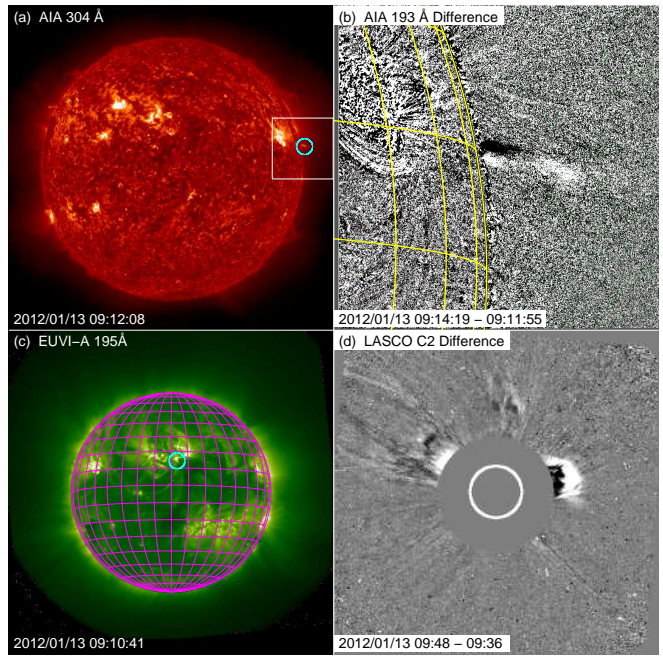}
\caption{Images for the $^{3}$He-rich event on 2012 January 13.  (a) AIA
  304~\AA\ images. The encircled area appears to contain an elongated
  structure. The box defines the field of view of the cutout image in
  (b).  
(b) AIA 193~\AA\ difference image barely revealing the jet.  (c)
  195~\AA\ image from EUVI on {\it STEREO-A} (107$\arcdeg$ west of the
  Sun-Earth line),
showing the brightening associated with the jet, and
  confirming its backside origin.  (d) LASCO C2 image showing the
  associated CME toward the northwest.}
\end{figure}

\begin{figure}
\includegraphics[scale=.45]{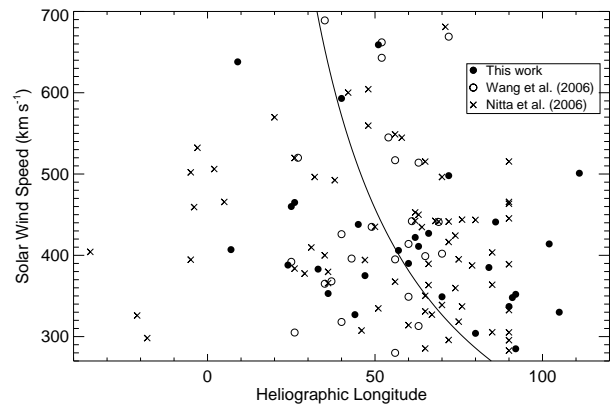}
\caption{The source regions are plotted in terms of the heliographic
  longitude and the associated solar wind speed.  The results from
  this work are compared with the earlier ones.  The curve shows the
  relation between the solar wind speed and the longitude of the nominal Parker spiral.}
\end{figure}

Next, we discuss where the $^{3}$He-rich events come from.
The location of the source region is shown in the seventh column of
Table~2.  The $^{3}$He-rich events occurred predominantly in the northern hemisphere up
to 2013, but all of the more recent events are in the southern
hemisphere.
This may simply reflect a larger number of active regions that emerged in the southern hemisphere during the period in question.
An important result comes from the direct observation of regions
behind the limb made possible by {\it STEREO}.
We find three cases where the source region is more than
10$\arcdeg$ behind the limb.  
An example is shown
in Figure~9 (Event 11).  This event is actually preceded by a less
intense $^{3}$He-rich period that lasts for $\sim$15 hours, but we study only the
later event, which is more intense. 
About five hours before the $^{3}$He onset, 
we find a strong type III burst, which is observed at the highest
frequency of WAVES even though it is limb-occulted as shown below. 
Around the time of the type III burst, we see a
diffuse linear feature stick out of the limb (Figure~9(b)).  Its
associated brightening is readily located in an EUVI-A image to be N16 W04
(Figure~9(c)).  Since {\it STEREO-A} was 107$\arcdeg$ west of the
Sun-Earth line, we determine the source location to be N16 W111 as
viewed from the Earth.
With reference to the type III burst, 
the delay of the electron event is only $\sim$12~minutes.  However, 
there is no velocity dispersion, and the time profiles are gradual.

The existence of such events naturally adds to the broad longitudinal
distribution of $^{3}$He-rich events, which was already found by 
\cite{2006ApJ...639..495W} and \cite{2006ApJ...650..438N}.
In Figure~10, we show a scatter plot of the longitude of the source
regions and the solar wind speed (the five hour average around the
type III burst as shown in the twelfth column), in comparison with the
earlier works. 
 We confirm that 
the longitudinal distribution is broad, often far from the longitude 
of the nominal Parker spiral, which is farther from the west limb for
faster solar wind as indicated by the curve in Figure~10.  In this work, examples of the source longitude
close to and behind the west limb are established for the first time
by direct observations without extrapolating the longitudes found earlier.

Now we compare the polarities of the source regions (the eighth
column) with those in-situ
at L1 around the times of $^{3}$He-rich events (the thirteenth
column).  The latter is determined with respect to the Parker spiral
for the observed solar wind speeds. We assign either positive (away)
or negative (toward) polarity as long as the azimuth angle of the
magnetic field measured at L1 is more than 15$\arcdeg$ 
from the normal to the Parker spiral.  In reality we
find that the observed field is often far from the 
Parker spiral \citep[cf.][]{1993JGR....98.5559L}.  The polarity of the
source region is less straightforward, since the region may not have a
dominant polarity.  We conduct PFSS extrapolations to find if the
region contains open field lines with the same polarity.  In several
regions we find no open field lines within a 20$\arcdeg$ heliographic 
distance, in which case we put C (closed) in the eighth column.  When open
field lines are identified in or near the source region, 
the polarities generally match at the Sun and L1.  

Following
\cite{2006ApJ...650..438N}, we also grade the performance of field
line tracing on the basis of the PFSS model and Parker spiral, respectively,
within and beyond the source surface.
We trace field lines from the
source surface to the photosphere.  Given the uncertainty in the
IMF, we include field lines that are within 
$\pm$7.5$\arcdeg$ and
$\pm$2.5$\arcdeg$, respectively, from the longitude and latitude of
the footpoint of the Parker spiral that intersects L1.  
We trace about 2000 field lines downward that are uniformly
  distributed in the above longitudinal and latitudinal ranges on the source surface.
Our grading is
1 (best) to 4 (worst) depending on the minimum
distance ($d_{min}$) of the footpoints of the traced field lines to the source
region: 1 if $d_{min} \lesssim 10\arcdeg$, 2 if $10\arcdeg < d_{min} \lesssim
20\arcdeg$,  3 if $20\arcdeg < d_{min} \lesssim 30\arcdeg$ and 4 
if $40\arcdeg < d_{min}$.  The results shown in the eighth column are
consistent with those by \cite{2006ApJ...650..438N}.  This simply
represents the status of how we model the Sun-Earth magnetic field
connection. 
The results may not improve drastically with
state-of-the-art numerical models rather than the simple PFSS $+$
Parker spiral model \citep{2011SpWea...910003M}, 
as long as the lower boundary conditions are set by the photospheric magnetic maps, only about 1/3 of which reflect direct observations.

\section{Discussion}

In this paper we use {\it SDO}/AIA data to identify the solar sources
of $^{3}$He-rich SEP events, which have been known to be weak in terms
of coronal signatures.  The significantly improved quality of AIA images do
reveal a much larger number of weak transients 
than did previous instruments (e.g. EIT).  Although it is possible
that EIT
missed many weak and short-lived transients, AIA data present 
another challenge of identifying the activity responsible for a given
$^{3}$He-rich SEP event from among the numerous other activities that
are occurring here and there all the time. 
Fortunately, we know that type III bursts are highly associated with
$^{3}$He-rich SEP events \citep{1986ApJ...308..902R,2006ApJ...650..438N}.
They let us narrow down the time range in
which we need to examine images, and it turns out to be relatively
easy to find the solar activity associated with the type III burst.
We also use the association of electron events as a
selection criterion because of their known correlation with
$^{3}$He-rich events \citep{1985ApJ...292..716R}.

We need to ask how valid is our approach that assumes that type III
bursts (and electron events to some extent) are a progenitor of
$^{3}$He-rich events.  Type III bursts often
accompany large gradual SEP events \citep{2002JGRA..107.1315C}.  
Many of them are better seen at
low frequencies, e.g. below 1 MHz, and delayed with respect to the
associated gradual flare.  But there are others that start at the
beginning of the associated impulsive flare and are seen at the
highest frequency of WAVES.  It is true that the appearances of type
III bursts and electron events alone do not distinguish between impulsive
($^{3}$He-rich) and gradual SEP events.  However,
the regions that produce large CMEs responsible for gradual events
tend to be different from the regions that produce smaller activities
responsible for  $^{3}$He-rich events.  It is usually trivial to
separate them in EUV images.

A lingering puzzle in the sources of $^{3}$He-rich events is the
height of the acceleration process in the corona.  
There is strong evidence that the
ionization state of the ions is caused by passage through coronal
material, and this requires a low coronal source 
\citep[e.g.][]{2006SSRv..123..217K, 2007ApJ...671..947K, 2006ApJ...645.1516D}.
However,
simple models have shown that the low energy ions in these events are
injected about 1 hour later than the electrons, suggesting a
high coronal source.  We have examined the five events in Table 1 
(Nos. 6, 9, 20, 25 and 26) that
have clear velocity dispersion such as shown in Figure 2, and 
found that fits to
the onset profiles yield ion injection times about 2 hours later than
the electron onset times shown in Table 2, roughly consistent with
earlier work.  Such simple fits assume that there is no scattering during
transport from the Sun, and this may not be the case for these ions.
\cite{2005ApJ...626.1131S} have discussed how interplanetary scattering can be
present and yet still produce a clear velocity dispersion pattern at 
1~AU where the path length traveled is not the IMF length 
but rather the IMF length divided by the average
pitch cosine for the particles.  Their model investigation showed how
this can produce inferred injection times 
significantly later than the actual injection time.  

In this study we find $^{3}$He-rich events produced by solar
eruptions that are not necessarily jets.  
However, none of these eruptions are 
very energetic CMEs like those that often accompany large gradual SEP events.
Furthermore, it is possible that the jet, even though it is present,
can be overlooked when the primary activity is a wider eruption.
Models have been developed to explain particle escape, which 
accommodate both jets and flux ropes \citep{2013ApJ...771...82M}. 
One of the surprises may be the involvement of EUV waves or LCPFs
(Figure~8(c)),
which used to be connected to large CMEs.
For larger, gradual SEP events
attempts have been made to reproduce the SEP onset behavior in terms of the
injection as the LCPF intersects the footpoint of the field line that
crosses the observer 
\citep[e.g.][]{1999ApJ...519..864K,2011ApJ...735....7R,2012ApJ...752...44R}.
Here we suggest the possibility that LCPFs may play a role in
$^{3}$He-rich events observed at widely separate longitudes
\citep{2013ApJ...762...54W} because several examples in Wiedenbeck et
al. also accompany LCPFs including the solar minimum event 
(on 2008 November 4) discussed
by \cite{2009ApJ...700L..56M} and our event 2. 
LCPFs may also lead to injection of particles in an open
  field region away from the flare site \citep{1999ApJ...519..864K}, which may be closed.
Again, we point out
that even those LCPFs could start off as jets.  The event shown in
Figure~8(c) is such an example.

Figure~10 shows the longitude and solar wind speed distributions for
the events in Table~1, along with previous recent studies by 
\cite{2006ApJ...639..495W} and \cite{2006ApJ...650..438N}.
The distribution is very wide,
covering more than the entire western hemisphere.  This contrasts with
the earlier view \citep{1999SSRv...90..413R}
that these events had western
hemisphere source locations peaked near $\sim$W60 that varied due to solar wind
speed, with some additional broadening from magnetic field line random
walk.  The more sensitive observations of recent years show that this is
not the case:  there is little evidence in Figure~10 of
the expected correlation between source longitude and solar wind
speed.  For the events in Table~2, the correlation coefficient between
source location and solar wind speed is -0.36.  
The p-value of 0.05 indicates that
this is a
statistically significant correlation for 29 events (including a
second injection in three of the 26 events), but it is clear
that solar wind speed does not dominate the variations in source
longitude.  Recent multi-spacecraft  studies have shown that some
$^{3}$He-rich events are observed over very wide longitude ranges
\citep{2013ApJ...762...54W}.

\cite{2012ApJ...751L..33G} have shown that perpendicular
transport combined with field line meandering may produce significant
transport of particles over wide longitudinal ranges, but it is not
clear if such models can reproduce the observed onset timing. These
problems combined with the limited success of the PFSS model in
predicting other properties such as field polarity or even connection
to known $^{3}$He-rich event sources \citep[e.g.][]{2009ApJ...700L..56M}
suggest that the difficulties encountered here are due 
at least partly
to an insufficiently
realistic model of the coronal magnetic field.  Recent models (S-web)
with more sophisticated approaches 
\citep[e.g.][]{2011ApJ...731..112A,2011ApJ...731..110L} 
regarding solar wind sources have shown that relatively
small regions on the Sun can be magnetically connected to large
regions in the inner heliosphere, as would be required to explain the
energetic particle observations discussed here.  These new models
require large computational resources so it is not a simple matter to
test them against a set of events such as studied here.  

Comparison of Tables~1 and 2 shows that $^{3}$He-rich events with
  high $^{3}$He/$^{4}$He ratios and curved $^{3}$He spectrum tend to
  be associated with weak of no GOES flares.  Moreover, they are more
frequently associated with jets than those that have power-law spectra.  However, the correlation is not strong, and we need a larger
sample to verify it.

\section{Conclusions}

Using high-cadence AIA images in combination with EUVI images for
areas not viewed from Earth, we have identified the source regions
of 26 $^{3}$He-rich SEP events in solar cycle 24, and classified the
activity at their source. Combined with prior work the basic
properties of these sources can be summarized as follows: 
\begin{enumerate}

\item We confirm the previously identified association of these events
  with type III radio bursts and electron events.
\item Solar activity at the source locations is generally weak
  in terms of soft X-ray flux, and
  can be associated with a variety of motions such as small eruptions,
  jets, and large-scale coronal propagating fronts.  
\item The broad longitudinal distribution of sources is not
  consistent with simple Parker spiral, nor with Parker spiral plus
  PFSS modeling of the coronal magnetic field.
  The actual magnetic connection between the sources
  and IMF may be much broader.
\item Besides these clear associations the correlation is not strong
  between the properties of the solar sources and  the energetic ions such as $^{3}$He/$^{4}$He ratio or spectral form.  
\end{enumerate}

Given the extremely high sensitivity of the AIA and ACE measurements,
it is likely that inner solar system measurements from upcoming
missions will be needed to settle many of the observational
ambiguities not resolved by the currently available data.

\acknowledgments
We thank the referee for letting us find and correct some
  problems in the original manuscript.
This work has been supported by the NSF grant AGS-1259549, NASA AIA
contract NNG04EA00C and the NASA STEREO mission under NRL Contract
No. N00173-02-C-2035.  GMM acknowledges NASA grant NNX10AT75G,
44A-1089749, and NSF grant 1156138/112111. CMSC and MEW acknowledge
support at Caltech and JPL from subcontract SA2715-26309 from UC
Berkeley under NASA contract NAS5-03131T, and by NASA grants
NNX11A075G and NNX13AH66G.









\clearpage

\end{document}